\documentstyle[twoside,fleqn,psfig,espcrc2,here]{article}
\newcommand{\be}{\begin{equation}}
\newcommand{\ee}{\end{equation}}
\newcommand{\bea}{\begin{eqnarray}}
\newcommand{\eea}{\end{eqnarray}}
\newcommand{\bean}{\begin{eqnarray*}}
\newcommand{\eean}{\end{eqnarray*}}
\newcommand{\ba}{\begin{array}}
\newcommand{\ea}{\end{array}}

\newcommand{\norsl}{\normalsize\sl}
\newcommand{\norsc}{\normalsize\sc}

\title{ \bf Electroweak Sudakov effect 
on processes at TeV scale 
\thanks{Talk at 
\norsl Theory Meeting on Physics
 at Linear Colliders, \normalfont  March 15 - 17, 2001, KEK, Japan.
This talk is based on work in collaboration 
with H. Kawamura and J. Kodaira~\cite{HKK,K}
}}

\author{
\norsc  Michihiro HORI 
\thanks{JSPS Research Fellow }
\thanks{E-mail address:michi@theo.phys.sci.hiroshima-u.ac.jp }\\
\vspace{3mm}
\norsl Dept. of Physics, Hiroshima University,
\norsl  Higashi-Hiroshima 739-8526, JAPAN
}
\begin{document}


\begin{abstract}
In Next Linear Colliders at TeV scale,  electroweak
double logarithmic corrections, which come from the infrared
behaviors of theory can not be neglected. It is well known that in QED and
QCD, double logarithmic corrections are resummed to all orders, 
and these corrections can be exponentiated, resulting 
in the Sudakov form factor. However it
is never trivial that double logarithmic corrections in
electroweak theory can be exponentiated, because of the spontaneous
 breaking of symmetry and the pattern 
of that. 
We discuss the electroweak double logarithmic corrections at two loop
 level and explain the differences of ``Soft'' structure between the
electroweak theory and QCD (the unbroken non-abelian gauge theory).
\end{abstract}

\maketitle
\section{Introduction}
Next Linear Colliders at TeV scale are planned for 
new physics search. 
These colliders are expected to have high luminosities 
and we will be able to perform accurate experiments.   
On the theoretical side, to extract new physics beyond Standard
Model (SM) from experimental data, higher order precision
calculations are required.\\ 
\hspace*{12pt}Recently, it is pointed out that 
logarithmic corrections in electroweak (EW)
theory are not negligible at TeV scale~\cite{CC1,CCC}. For example,
EW 1-loop corrections for the process 
$ e^{+}e^{-}\rightarrow \mu^{+}\mu^{-}$ 
are discussed~\cite{BCCRV}. The logarithmic corrections for this process 
dominate the cross section when total energy goes up to TeV region. These 
logarithmic corrections are classified into the ultraviolet (UV)
logarithm, the single infrared (IR) logarithm which is the contribution from 
soft or collinear region in the loop integration, and the Sudakov type
double logarithm (DL)~\cite{SU} which is originated from soft 
and collinear region in
the loop integration. Particularly, Sudakov type DL
correction is of order 10\% at $\sqrt{s}=$1 TeV. These DL
corrections may spoil the perturbative prescription when total energy
grows up beyond TeV scale. Therefore we
must control the DL corrections to obtain the reliable predictions. 
In QED and QCD, it is well known that we can resum 
the DL corrections to all orders, resulting in the Sudakov 
form factor~\cite{JCC}.   
However, the EW theory is more complicated than QCD (unbroken
non-Abelian gauge theory) in two aspects. Firstly, the symmetry is
spontaneously broken. Secondly, the pattern of the symmetry breaking is
that the off-diagonal $U(1)_{em}$ part of $ SU(2)\otimes U(1)$ is survived. 
These lead to the mass difference between the gauge bosons
 and the mixing of neutral gauge bosons. Therefore it is 
non-trivial that the EW DL corrections can be exponentiated. 
If we can't control these large DL corrections, 
the perturbative approach can not be trusted beyond TeV scale
in EW theory.\\
\hspace*{12pt} This problem has been discussed to all orders by several
authors~\cite{CC2,KP,FLMM}. K\"uhn and Penin~\cite{KP}
considered the process $e^{+}e^{-}\rightarrow f\bar{f}$ in the Coulomb
gauge. They conclude that EW DL corrections 
are not exponentiated. But they have taken into account only W and Z
contributions to the process. Ciafaloni and Comelli~\cite{CC2}
considered the process $Z'\rightarrow f\bar{f}$ in the Feynman gauge.
They use the Soft insertion formula which has been developed in 
QCD~\cite{C}. They assume the ``strong energy ordering'' to gauge bosons
attached to a fermion line with Eikonal current, namely the energies 
of external boson lines are 
smaller than the ones of inner boson lines in the diagram. They conclude
that the EW DL corrections cannot be
exponentiated. However this method can take into account only ladder 
diagrams. Fadin et al.~\cite{FLMM} discuss on the exponentiation using
the IR evolution equation for the amplitude which
is a function of the infrared cut-off, and this equation is analogous to the
renormalization group equation. They conclude that EW DL corrections 
can be exponentiated. These papers disagree with each
others. \\
\hspace*{12pt}In order to solve this controversy, 
several authors calculated explicit 2-loop DL
corrections~\cite{BW,M2,HKK}. Beenakker and Werthenbach~\cite{BW}
consider the process $e^{+}e^{-}\rightarrow f\bar{f}$ in the Coulomb
gauge. Melles~\cite{M2} consider the process $g \rightarrow
f_{R}\bar{f_{L}}$ in the Feynman gauge. Since only photon and Z boson
contribute to this process, we want to consider the general case
including W boson contribution. We consider 
the process $g \rightarrow f_{L}\bar{f_{R}}$ in the Feynman gauge. 
In the next section, we show whether the exponentiation of EW Sudakov
type DL corrections holds at the 2-loop level in this process, 
and discuss the difference 
of ``Soft'' structure between the EW theory and QCD (the
unbroken non-abelian gauge theory).
\section{Explicit calculation of DL corrections}
In this section, we give an explicit 2-loop calculation of DL
corrections to the  fermion's form factor in the 
Feynman gauge. The masses of W and Z bosons will be approximated to 
be equal $M_{W}\simeq M_{Z}
\equiv M$~\footnote{Because we consider the leading double logarithmic
contribution, we need not take into account the mass difference 
between W and Z boson.}. We give a fictitious small mass to
photon to regularize the IR divergence and fermion is assumed 
to be massless. We
consider the situation, $s \gg M \gg \lambda$, where s is the total energy of 
produced fermions. In section 2.1, we review DL
corrections in QCD. In 2.2, we estimate DL corrections
in EW theory. And we discuss the difference and the
similarity of IR structure between EW theory and QCD.    
\subsection{DL corrections in QCD}
There are many investigations of 
QCD DL corrections to the fermion's form factor~\cite{FFT}. 
The QCD 1-loop DL contribution is, \\
\vspace*{-1cm}
\begin{figure}[H]
\psfig{file=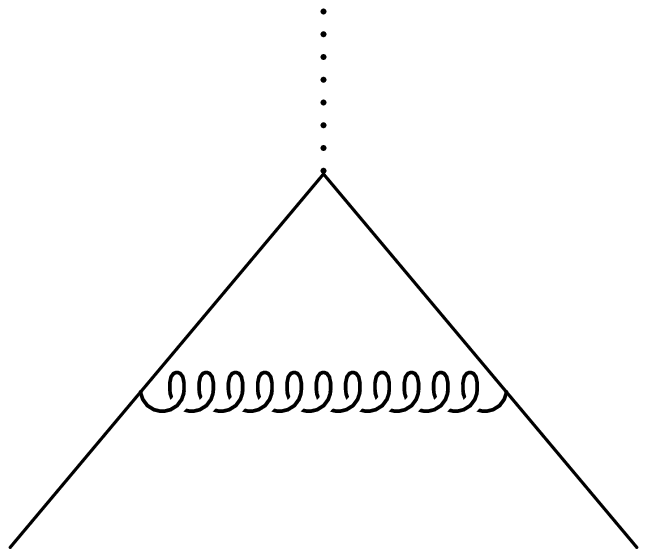,width=2cm}
\vspace{-1cm}
${\hspace*{3cm}\mbox{\Large:}\hspace*{0.5cm} -\frac{g_{s}^{2}}{16\pi^{2}}\ln^{2}\frac{s}{\lambda^{2}}C_{F}}$,\\
\vspace{-0.6cm}
\end{figure}
where $C_{F}$ is the SU(3) Casimir operator for fundamental
representation and $g_{s}$ is the strong coupling constant.\\
\hspace*{12pt}Next, we consider the 2-loop DL contribution.
The diagrams which contribute at 2-loop level are,\\    
\vspace*{-1cm}
\begin{figure}[H]
\psfig{file=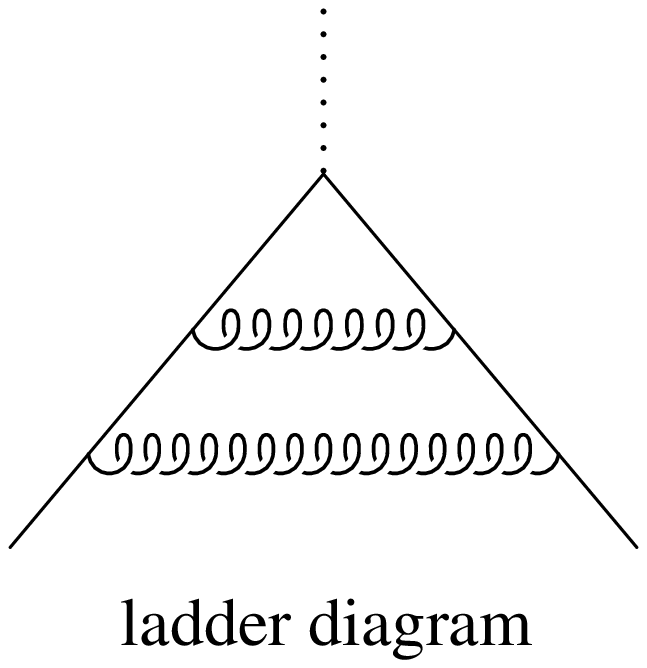,height=2cm}
\vspace{-1.3cm}
$\hspace*{2.5cm}\mbox{\Large:}\hspace*{0.5cm} \frac{g_{s}^{4}}{(8\pi^{2})^{2}}\frac{1}{24}\ln^{4}\frac{s}{\lambda^{2}}C_{F}^{2}$,\\
\vspace{-0.6cm}
\end{figure}
\vspace*{-1cm}
\begin{figure}[H]
\psfig{file=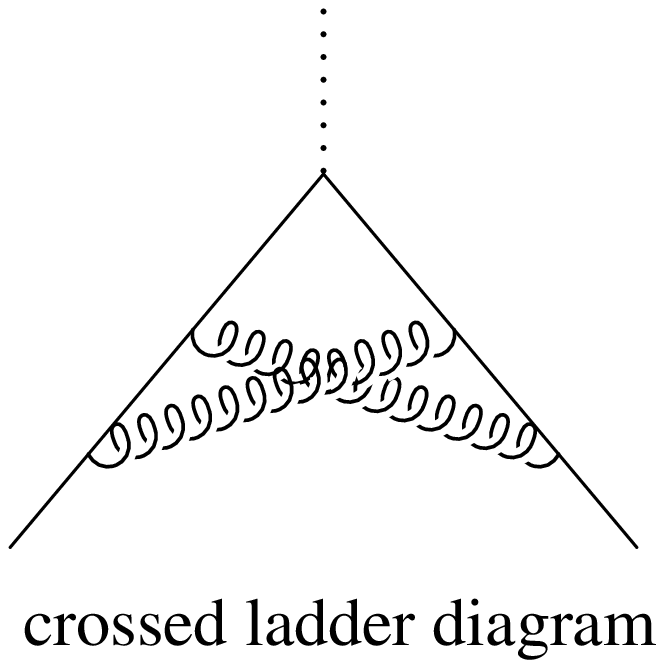,height=2cm}
\vspace{-1.3cm}
$\hspace*{2.3cm}\mbox{\Large:}\hspace*{0.5cm} \frac{g_{s}^{4}}{(8\pi^{2})^2}\frac{1}{12}\ln^{4}\frac{s}{\lambda^{2}}(C_{F}^{2}-\frac{1}{2}C_{A}C_{F})$,\\
\vspace{-0.6cm}
\end{figure}
\vspace*{-1cm}
\begin{figure}[H]
\psfig{file=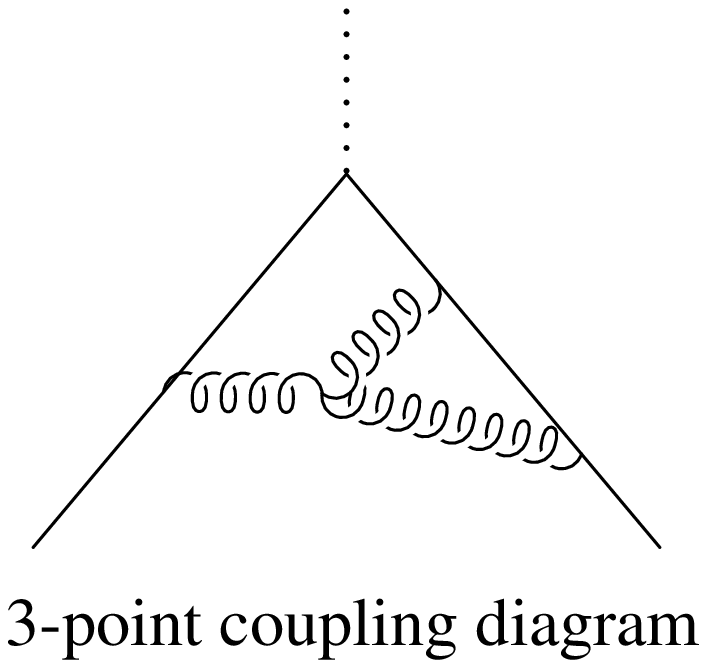,height=2cm}
\vspace{-1.3cm}
$\hspace*{2cm}\times 2\hspace*{0.5cm}\mbox{\Large:}\hspace*{0.5cm}\frac{g_{s}^{4}}{(8\pi^{2})^2}\frac{1}{12}\ln^{4}\frac{s}{\lambda^{2}}\frac{1}{2}C_{A}C_{F}$,\\
\vspace{-0.6cm}
\end{figure}
where $C_{A}$ is the Casimir Operator for adjoint representation. The
factor 2 in the contribution of diagrams which have 
the triple gauge boson coupling
comes from the symmetric diagram. Note that 
the second term in the crossed ladder diagram appears as a result of the
non-abelian nature of SU(3). But, this term is cancelled out by the
contribution of the 3-point coupling diagrams.
Therefore, the 2-loop contribution becomes $\frac{1}{2}$(1-loop
contribution)$^{2}$, and we find that the exponentiation of 
QCD DL corrections holds at 2-loop level. \\
\subsection{DL corrections in EW theory}
We devote this section to discussion of the EW DL
corrections. We consider the process of the production of 
the left handed fermion and the right handed antifermion from a
$SU(2)\otimes U(1)$ singlet source. \\
\hspace*{12pt}First, we consider 1-loop DL contribution.
The diagrams which contribute at 1-loop level are,\\
\psfig{file=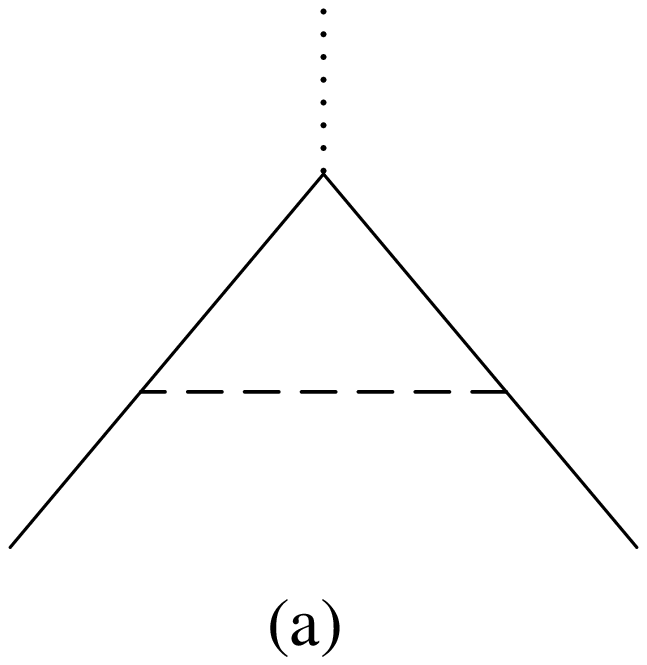,height=3cm}
\hspace{1cm}
\psfig{file=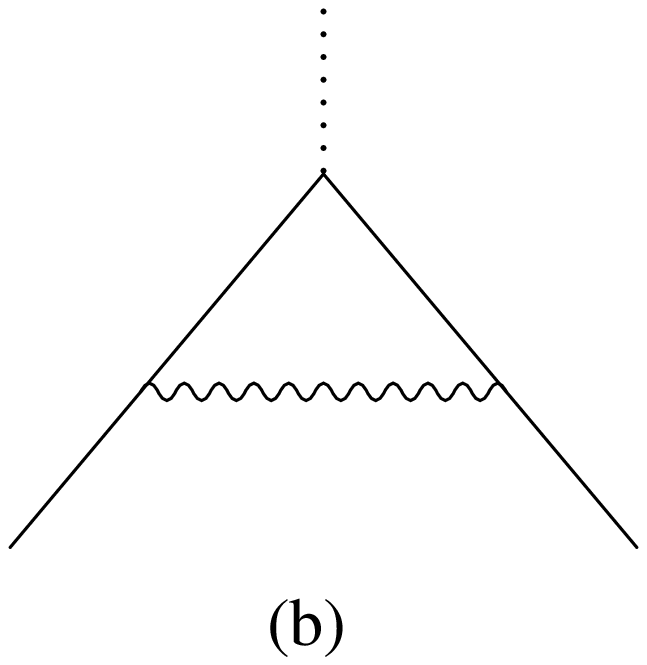,height=3cm}\\
, where the dashed line is photon, the wavy line is W or Z boson. 
We present the
group factor of SU(2)$\otimes$U(1) and the kinematical factor of loop
integration separately. The group factors become,
\bean
    \gamma\ {\rm exchange}\ &:& e^2 Q^2 \ ,\\
     W \ {\rm exchange}\ &:& g^2 \sum_{a= 1,2} T^a T^a \ ,\\
     Z \ {\rm exchange}\ &:& g^2 T^3 T^3 + g'^2 Y^2 - e^2 Q^2 \ ,
\eean
where $Q = T^{3} + Y$, $T^{a}$ are the SU(2) generators, Y is the 
hypercharge, g and g' are SU(2) and U(1) coupling constants and e is the 
electric charge. The loop integrals in which gauge bosons are exchanged produce
the following double logarithms.
\bean
{\rm(a)\hspace*{5pt}diagram}\ &:& -\frac{1}{16\pi^{2}}\ln^{2}\frac{s}{\lambda^{2}} \\
{\rm(b)\hspace*{5pt}diagram} \ &:& -\frac{1}{16\pi^{2}}\ln^{2}\frac{s}{M^{2}}. \\
\eean
By combining the group factor and the loop kinematical factor, the
result of form factor up to 1-loop is, \\
\bean
    \Gamma^{(1)} =  1 
   &-& \frac{1}{16 \pi^2} \left( g^2 C_{F} + g'^2 Y^2 - e^2 Q^2 \right)
              \ln^2 \frac{s}{M^2} \\
    &-& \frac{1}{16 \pi^2} e^2 Q^2
                  \ln^2 \frac{s}{\lambda^2} \ ,
\eean
where $C_{F}$ is the SU(2) Casimir operator for the fundamental
representation.\\
\hspace*{12pt}Next, we consider 2-loop DL
contributions. The diagrams which contribute at 2-loop level are as
follows.\\
$\bullet$ ladder diagrams \\
\psfig{file=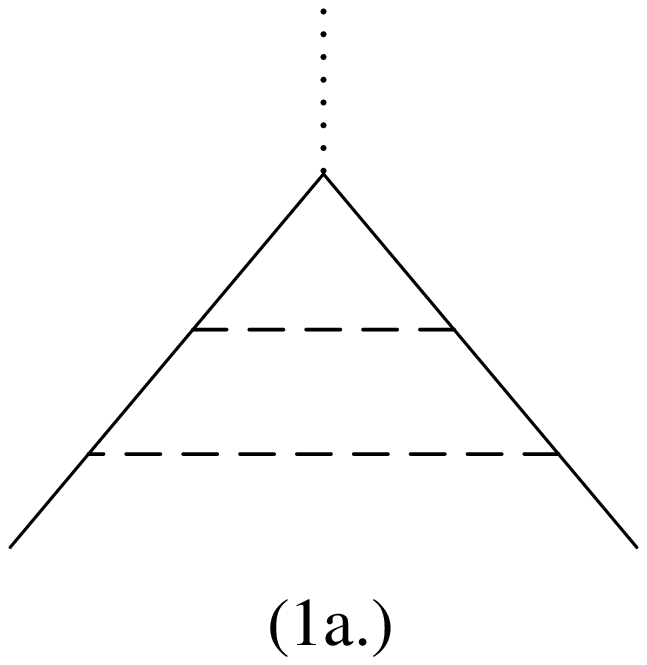,height=2.5cm} 
\psfig{file=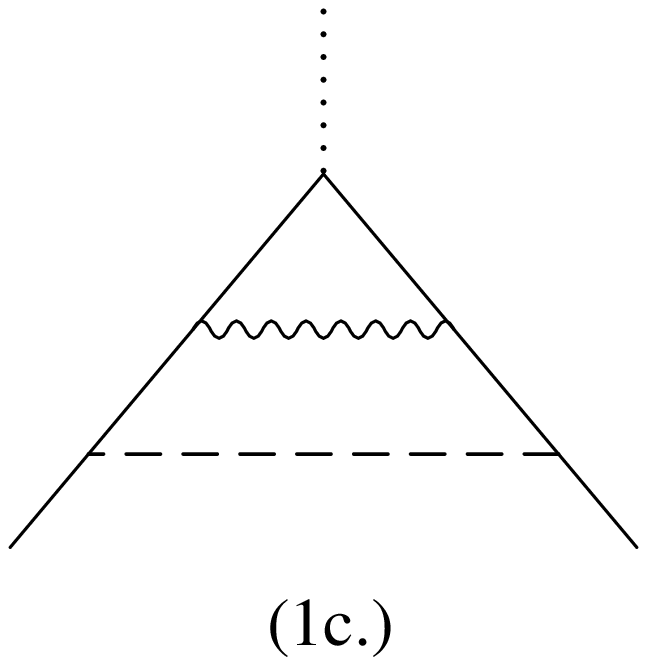,height=2.5cm} 
\psfig{file=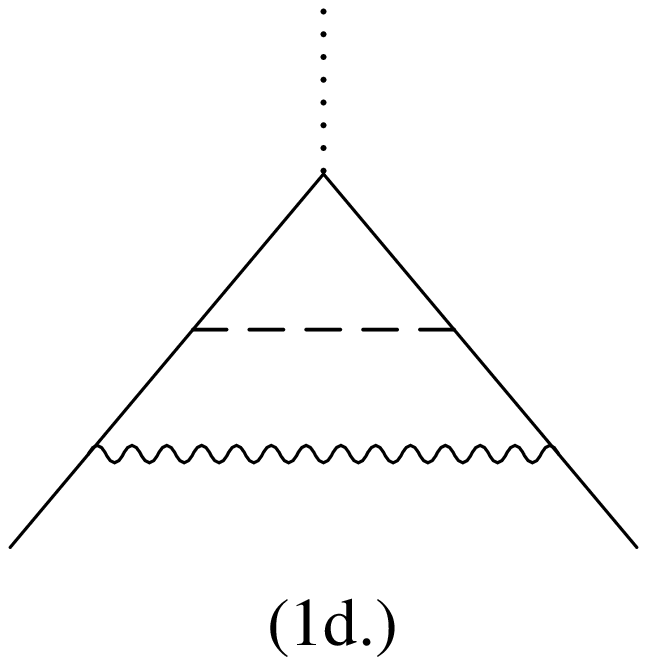,height=2.5cm} 
\hspace*{2.5cm}\psfig{file=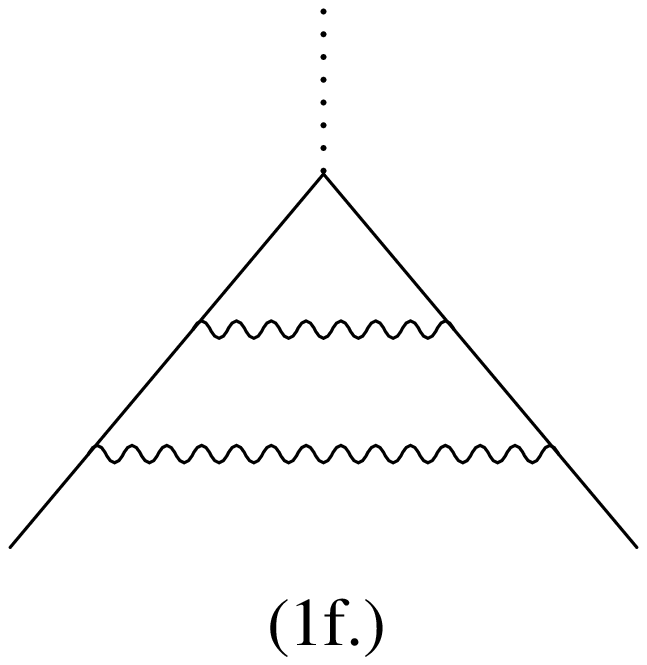,height=2.5cm} \\
$\bullet$ crossed ladder diagrams \\
\psfig{file=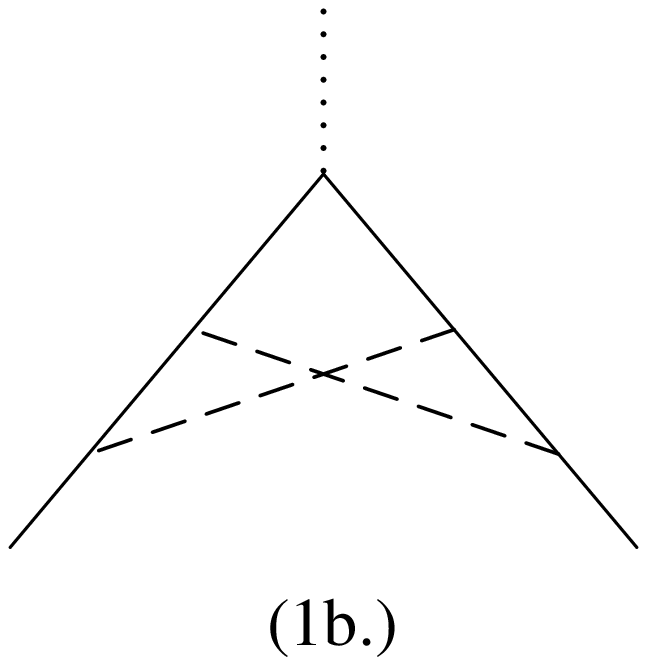,height=2.5cm} 
\psfig{file=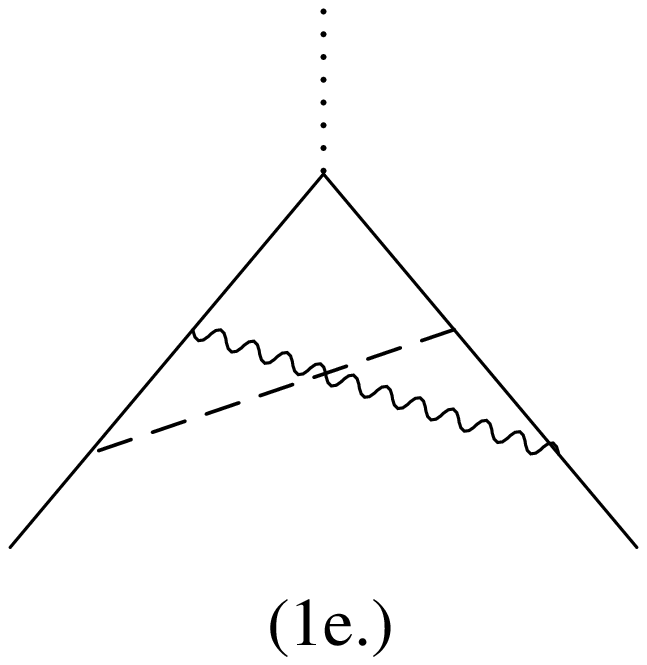,height=2.5cm} 
\psfig{file=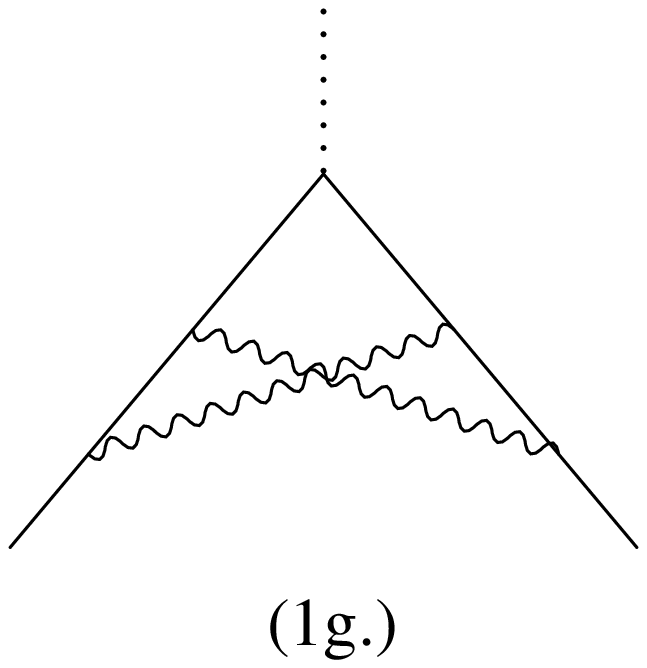,height=2.5cm} \\
$\bullet$ 3-point coupling diagrams \\
\psfig{file=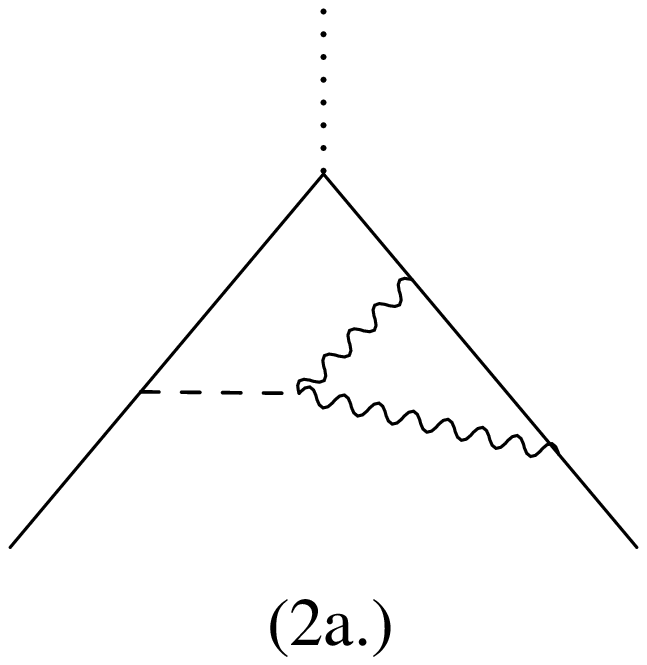,height=2.5cm} 
\psfig{file=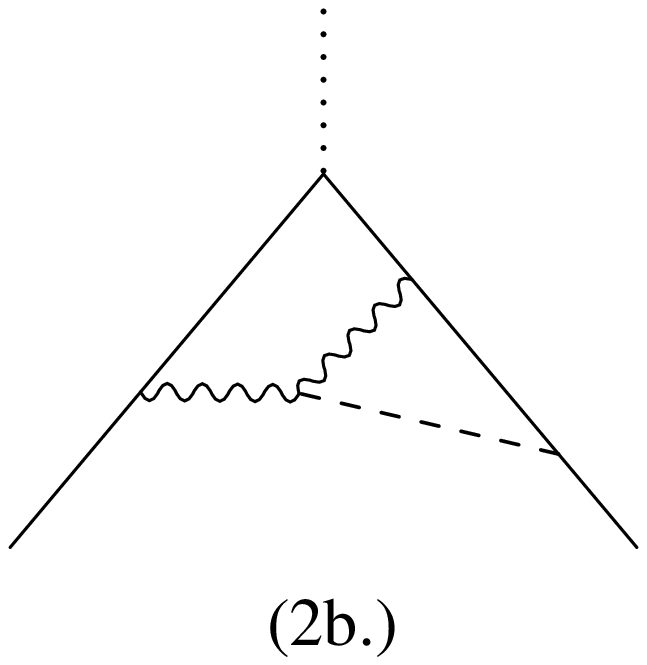,height=2.5cm} 
\psfig{file=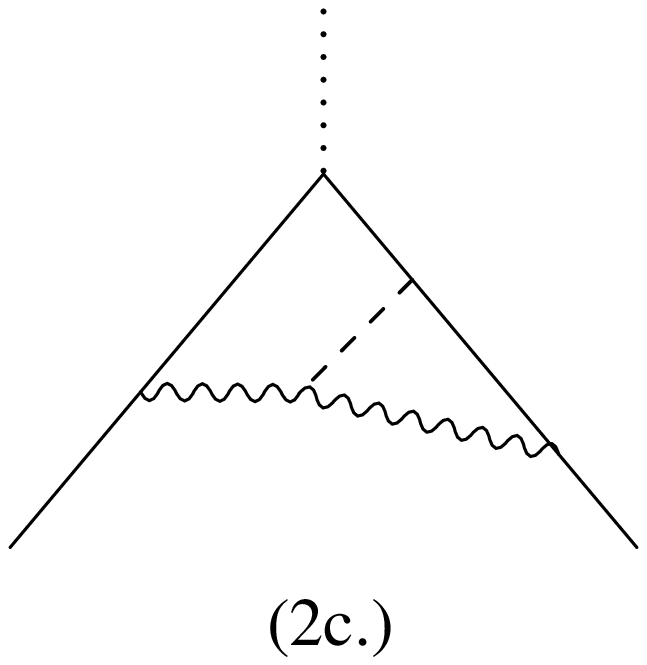,height=2.5cm} 
\hspace*{2.5cm}\psfig{file=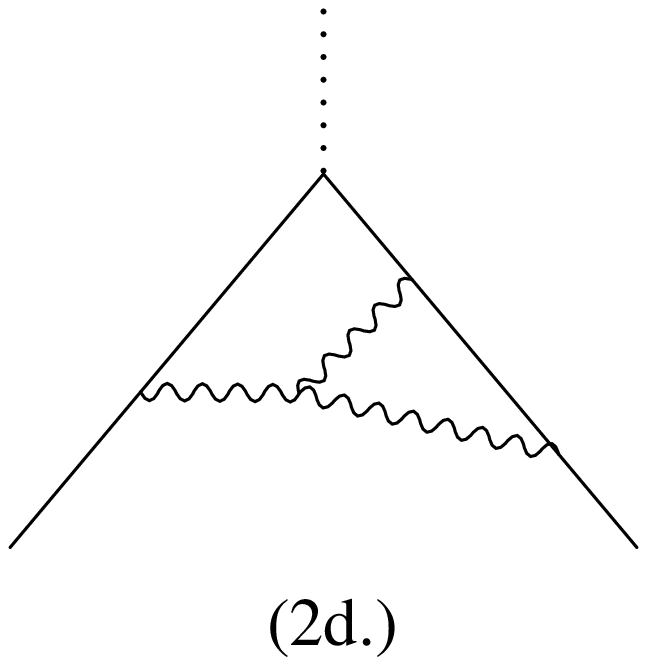,height=2.5cm} \\
\hspace*{12pt}For notational simplicity, 
group factors and loop integral factors are 
written as,\\
\bean
    {\rm group\hspace*{5pt} factor}\ &:& \gamma\equiv e^{2}Q^{2}, \\  
           & &  W + Z \equiv g^2 C_{F}+ g'^2 Y^2 - e^2 Q^2.\\
 {\rm loop\hspace*{5pt}integral\hspace*{5pt}factor}\ &:& \l\equiv\ln\frac{s}{\lambda^{2}},\hspace{0.5cm}L\equiv\ln\frac{s}{M^{2}}.\\
\eean 
The contribution of each diagram reads. 
\bean
&&(\mbox{ladder contribution}) \\
& &= 
\frac{1}{(8\pi^{2})^{2}}\left[\gamma^{2}\frac{1}{24}l^{4}\right.\\
&&\left.+ 2\gamma\left(W + Z\right)\frac{1}{8}L^{2}l^{2} 
+ \left(W + Z\right)^{2}\frac{1}{24}L^{4}\right.\\
&&\left.+\gamma\left( W + Z \right)\left\{\frac{1}{6}L^{4}
-\frac{1}{3}L^{3}l\right\}\right] ,\\
&&(\mbox{crossed ladder contribution}) \\
& &=
\frac{1}{(8\pi^{2})^{2}}\left[\gamma^{2}\frac{1}{12}l^{4}\right. \\
&&\left.+\left\{\left(W + Z\right)^2 -
g^{4}\frac{1}{2}C_{A}C_{F}\right\}\frac{1}{12}L^{4} \right.\\
&&\left. 
+ \gamma\left(W +
Z\right)\left(-\frac{1}{6}L^{4}+\frac{1}{3}L^{3}l\right)\right.\\
&&\left.
+ 2 g^{2} e^{2} Q T^{3}\left(\frac{1}{6}L^{4}-\frac{1}{6}L^{3}l\right)
\right], \\
&&(\mbox{3-point coupling contribution})\\
& &=\frac{1}{(8\pi^2)^2}\left[g^4\frac{1}{2}C_{A}C_{F}\frac{1}{12}L^{4}
\right.\\
&&\left.
+2 g^2 e^2 Q T^{3} \left\{-\frac{1}{6}L^4 +
\frac{1}{6}L^3 l\right\}\right],\\
\eean
where $C_{A}$ is the SU(2) Casimir operator for the adjoint
representation. We find that
even the ladder diagram contribution 
has an non-exponentiating term(4-th term) which does not 
emerge in QCD. But this term is cancelled out by the third
term in the crossed ladder contribution. And the term proportional to
the Casimir operator for adjoint representation appears in the
crossed ladder contribution due to the non-abelian nature of
SU(2). This term is cancelled out by the first term in the 3-point coupling
contribution as in QCD. Other non-exponentiating terms(the 4-th term
in the crossed ladder contribution and the second term in the 3-point coupling
contribution) cancel each other. We have shown that non-exponentiating
terms are
cancelled out completely, and obtain the exponentiation of Sudakov form
factor at 2-loop level as follows.
\\
\\
\bean
    \Gamma^{(2)} &=&  1 - 
           \frac{1}{16 \pi^2} \left\{\left(W+Z\right)
              L^{2} + \gamma\l^{2}\right\}  \\
           && + \frac{1}{2!}
            \left[\frac{1}{16 \pi^2} 
       \left\{\left(W+Z\right)L^{2} + \gamma l^{2}\right\}\right]^2 \ .
\eean   
\section{Summary and Conclusion}
We have considered the electroweak form factor at 2-loop level in the DL 
approximation. We have used the standard Feynman gauge. Our
results have shown the exponentiation of the EW Sudakov form
factor at 2-loop level like QED and QCD. \\
\hspace*{12pt}This results are very important for theoretical  
predictions because these support the validity of 
the perturvative approach in EW theory beyond 
TeV scale. And this EW Sudakov effect
have to be taken into account on processes at TeV scale in future
colliders to obtain reliable predictions.
\section*{Acknowledgments}

The author would like to thank H. Kawamura and J. Kodaira for fruitful
discussions through this collaboration.
The work of M. H. was supported by the Grant-in-Aid for 
Scientific Research from the
Ministry of Education, Culture, Sports, Science and Technology, Japan,
No.11005244.


\end{document}